\documentstyle[11pt,aaspp4]{article}
\slugcomment{To be appeared in The Astrophysical Journal April 20, 1999
issue (Vol. 515)}
\begin{document}
\title{The Radio Emission of the Seyfert Galaxy NGC~7319
\footnote{Based on observations made with the Very Large Array 
operated by the National Radio Astronomy Observatory and on observations 
made with the NASA/ESA Hubble Space Telescope, 
obtained from the data archive at the Space Telescope Science Institute, which 
is operated by the Association of Universities for Research in Astronomy, Inc., 
under NASA contract NAS 5-26555.}}
\author{Kentaro Aoki}
\affil{Astronomical Data Analysis Center, 
National Astronomical Observatory of Japan, 
\\2-21-1, Osawa, Mitaka, Tokyo 181-8588 Japan}
\author{George Kosugi}
\affil{Subaru Telescope, National Astronomical Observatory of Japan,
\\650 North A'ohoku Place, Hilo, HI 96720}
\author{Andrew S. Wilson\altaffilmark{2}}
\affil{Astronomy Department, University of Maryland, College Park, MD 20742-2421}
\and
\author{Michitoshi Yoshida}
\affil{Okayama Astrophysical Observatory, National Astronomical Observatory of Japan,
\\Kamogata-cho, Asakuchi-gun, Okayama 719-0232, Japan}

\altaffiltext{2}{Adjunct Astronomer, Space Telescope Science Institute}

\begin{abstract}
We present VLA maps of the Seyfert 2 galaxy NGC~7319 at 3.6, 6, and 20 cm.
Sub-arcsecond resolution is achieved at 3.6 and 6 cm. 
The radio emission exhibits a triple structure on a scale of $\sim4$\arcsec\ 
(1.7 kpc).
All three components have steep spectra,
consistent with synchrotron radiation.
We have also analyzed an {\it HST} archival, broad-band red image,
which contains structure related to the radio components.
In particular, a V-shaped feature in the HST image some
3\farcs7 (1.6 kpc) southwest of the nucleus 
is associated with highly blueshifted emission lines seen in 
ground-based spectra.
We interpret the V-shaped feature as emission from gas compressed by
the bow shock driven by the outwardly moving radio plasmoid.
\end{abstract}
\section{INTRODUCTION}
A number of investigations have been made of the
relationship between the radio emission and 
the narrow-line regions (NLR) of Seyfert galaxies.
From ground-based observations, it is found that the NLR are aligned
and cospatial with the radio emission 
(Haniff, Wilson, \& Ward 1988).
There is also evidence that the kinematics of the NLR are 
related to the radio-emitting plasma.
Whittle et al. (1988) studied [\ion{O}{3}] $\lambda5007$ emission-line profiles
in several Seyfert galaxies which have linear
(i.e. double, triple or jet-like) radio sources.
They found that substructures (subpeaks and shoulders)
in the line profiles are more conspicuous 
close to the radio components.
This result suggests
that the radio-emitting plasma interacts with the optical line-emitting gas.

\par
The {\it Hubble Space Telescope} ({\it HST}) has allowed much higher
resolution imaging of the NLR and several studies
have focussed on the relationship
between the NLR and the radio emission 
(Bower et al. 1994, 1995; Capetti et al. 1996;
Capetti, Macchetto \& Lattanzi 1997; 
Falcke, Wilson \& Simpson 1998).
These observations have shown detailed correspondences 
between optical emission-line
and radio continuum images, and indicate that
interactions between radio plasma and the ambient gas determine 
the morphology of the NLR, with the radio ejecta
sweeping up and compressing the interstellar medium.
Detailed observations of such 
interactions are, therefore, necessary to understand the NLR within 
several hundreds pc of the nucleus. 
\par
NGC~7319, a member of Stephan's Quintet,
is a type 2 Seyfert galaxy with circumnuclear, outflowing, ionized gas
which aligns with the radio emission (Aoki et al. 1996).
Multiple components are found in the optical emission-line profiles
on the SW side of the nucleus.
The velocity of the outflow ranges 
up to 500 km s$^{-1}$ and its extent is 4 kpc.
NGC 7319 thus exhibits one of the largest circumnuclear outflows known
in Seyfert galaxies.
\par
In this paper, we report
sub-arcsecond resolution
radio imaging of NGC~7319 with the Very Large Array (VLA).
These radio images are
compared to an {\it HST} archival broad-band WFPC2 image
and the results of ground-based optical spectroscopy by Aoki et al. (1996)
in order to study in detail the relationship
between the radio emission and the outflowing gas.
Our VLA observations and the HST image are described in Section 2.
The results are presented in Section 3, while in Section 4 we discuss 
the properties of the radio-emitting
plasma in NGC~7319 and interpret our results.
We give concluding remarks in Section 5.
The heliocentric systemic velocity
of 6740 km s$^{-1}$
(Aoki et al. 1996) gives a distance of 86 Mpc for
NGC~7319 assuming a Hubble constant $H_{0}$=75 km s$^{-1}$ Mpc$^{-1}$ 
and the solar motion relative to the CMB radiation field (Smoot et al. 1991). 
Thus 1\arcsec\ corresponds to 420 pc.

\section{OBSERVATIONS AND REDUCTION}
\subsection{VLA Data}
The VLA observations of NGC~7319 were made in the `A-configuration' 
at 20, 6 and 3.6 cm on 1996 November 4.
The integration times on NGC~7319 were 1.4, 1.3, and 1.2 hours 
at 20, 6, and 3.6 cm, respectively.
Each 6 and 3.6 cm band consisted of two contiguous channels with a total
bandwidth of 100 MHz centered at 4860 and 8440 MHz, respectively. 
Two separate 50 MHz channels centered at 1365 and 1435 MHz were 
used for the 20 cm observations.
The data were phase calibrated 
by means of observations of the source 2236+284,
which is 6$^{\prime}$ away from NGC~7319.
The calibrator and NGC~7319 were observed sequentially with a 15 min cycle.
Flux calibration was achieved by observations of 3C48. 
\par
Data reduction was done with AIPS in Socorro.
The observations 
of NGC~7319 were flux and phase calibrated, mapped, and 
CLEANed to give the final images.
Due to an error, we observed 3C48 at 1385 and 1465 MHz, which are
different from the 20 cm bands used for NGC 7319 and 2236+284.
The fluxes of 2236+284 at its observed frequencies of 1365 and 1435 MHz
were therefore determined by interpolation and extrapolation of the
3C 48 observations, and used to flux calibrate the observations of
NGC 7319.

\subsection{{\it HST} Data}
There is a broad-band image of NGC~7319 in the {\it HST} archive.
It was obtained with WFPC2 
through the filter F606W (effective wavelength/width 5843~\AA/1578.7~\AA).
The nuclear region of NGC~7319 was imaged
with the Planetary Camera,
which has a pixel size of 0\farcs0455.
Bias and dark subtraction, and flat fielding
were done with pipeline processing at 
the Space Telescope Science Institute.
We removed cosmic ray events from the data using IRAF
\footnote{IRAF is distributed by the National Optical Astronomy 
Observatories, which are operated by the Association of Universities for
Research in Astronomy, Inc. under cooperative agreement with the National Science 
Foundation.}.
Note that the F606W filter contains strong emission-lines, 
such as
[\ion{O}{3}] $\lambda 5007$, H$\alpha$, and [\ion{N}{2}] $\lambda 6584$,
from NGC~7319.
The system throughputs
at [\ion{O}{3}] $\lambda 5007$ and H$\alpha$ are 60 \% and 90 \% 
of the peak, respectively, so the structures in this image are influenced by 
emission-lines.
\section{RESULTS}
\subsection{Radio Images}
The full resolution 3.6, 6, and 20 cm maps are presented in 
Figure 1, 2, and 3, respectively.
Three compact components, plus diffuse emission, are visible in the
higher resolution maps.
The positions, flux densities, and sizes (FWHM) of the
compact components were measured 
using the AIPS task JMFIT, which is a 
two-dimensional elliptical gaussian fitting program. 
These parameters are summarized in Table 1.
The sizes are given after deconvolution from the beams. \par
In the highest resolution ($0\farcs27\times0\farcs26$~FWHM) 
map (Fig. 1), the three compact components, labeled A, B and C,
are found to be aligned in P.A. $24\pm1.4$\arcdeg.
The separations of A and B, B and C, A and C are $0\farcs97$ (410 pc), 
$3\farcs36$ (1.4 kpc), and $4\farcs32$ (1.8 kpc), respectively.
Component A has a size of $0\farcs3 \times 0\farcs2$ (130 pc $\times$ 80 pc) 
and is elongated in P.A.$\sim 0 \arcdeg$.
Component C has a size of $0\farcs25 \times 0\farcs2$ (110 pc $\times$ 80 pc)
and is elongated in P.A.$\sim 40 \arcdeg$.
There is a diffuse component which extends $\sim 2\arcsec$ towards
the north from C.
There is also diffuse emission extending to the south of A.

\par 
The structure in the 6 cm naturally weighted map 
($0\farcs47\times 0\farcs45$~FWHM) (Fig. 2)
is similar to that in the 3.6 cm map. 
The three components (A, B, and C) are clearly seen, 
with diffuse emission extending between A and C.
The overall appearance of two outer `hot spots' (A and C)
with diffuse emission extending back towards the nucleus is
reminiscent of FRII-class radio
galaxies, though with much lower radio power and smaller spatial extent
in NGC 7319.

\par
The A and B components which are seen in the 3.6 cm and 6 cm images
are merged together
in the 20 cm uniformly weighted image 
($1\farcs31\times 1\farcs29$~FWHM) (Fig. 3).
The position of this merged
peak agrees with that given by van der Hulst \& Rots (1981) to
within 0\farcs2.
The resolution of our 20 cm image is higher than that of
van der Hulst \& Rots (1981), and the 
jet-like feature seen in their Figure 3 
corresponds to our component C.
Our 20 cm total flux density of $28.5\pm0.5$ mJy agrees with that found by
van der Hulst \& Rots (1981).
\par
The spectral indices ($\alpha$: $f_{\nu}\propto \nu^{\alpha}$) of the three 
compact
components were calculated from the 3.6 and 6 cm
flux densities and are given in Table 1.
The spectral index between 20 and 6 cm is also derived for component C.
The spectral indices of all three components are similar
($\alpha \sim -1.3$).
Our derived spectra for components with diffuse emission may be 
systematically too steep as the shorter wavelength maps may miss extended flux.
However, the steep spectra are consistent with the spectral index of the
total radio emission ($\alpha = -1.2$, van der Hulst \& Rots (1981);
$\alpha = -1.1 \pm 0.3$, 
Kaftan-Kassim, M. A., Sulentic, J. W., \& Sistal, G. (1975)).

\subsection{Optical Image}
The {\it HST} F606W image of NGC~7319 is shown in Figure 4.
Prominent spiral-like structure is seen around the nucleus.
The spiral features extend
1\farcs4 (590 pc) to the north and 1\arcsec (420 pc) to the south
of the nucleus.
The northern arm is associated with a dust lane. 
There is also a V-shaped feature 3\farcs4 south and 1\farcs4 west
of the nucleus.
To enhance the contrast of the fine-scale structure,
we  smoothed the image with a gaussian function of 1\farcs5 FWHM 
and subtracted it from the original (unsmoothed) one.
The difference image is shown in Figure 5. 
The V-shaped feature, as well as the spiral-like structure mentioned above,
are clearly seen.

\par
Both the spiral-like
structure and the V-shaped feature may, at least in part, be
emission-line regions,
since strong emission-lines ([\ion{O}{3}] $\lambda 5007$, H$\alpha$, and 
[\ion{N}{2}] $\lambda 6584$) are present in the filter passband.
We have estimated the contribution of emission-lines to
the spiral-like structure and the V-shaped feature 
in the F606W image using ground-based spectra of the corresponding region.
Nuclear (aperture size 4\farcs4 in NS $\times$ 1\farcs8 in EW) and off-nuclear
(aperture size 1\farcs8 in NS $\times$ 4\farcs4 in EW)
spectra were extracted.
The two extraction windows include the spiral-like structure
and the V-shaped feature, respectively.
These spectra were multiplied by the transmissions of 
{\it HST}, WFPC2 and F606W using the synthetic
photometry package SYNPHOT in STSDAS.
The result is that emission lines contribute
17 \% and 3 \% of the total flux of the nuclear region and off-nuclear region,
respectively.
We also derived the fluxes in the HST original (Fig. 4)
and difference (Fig. 5) images within the same apertures
as used to extract the ground-based spectra.
The spiral-like structure (as measured from the difference image)
comprises 14\% of the total flux (from the original image) through the
nuclear aperture. The corresponding number for the V-shaped
feature through the off-nuclear aperture is 3\%.
These fractions are similar to the contributions of emission-lines
in each aperture, so
it is possible that both structures are emission-line features with the 
smoothly distributed background being stellar light.

\section{DISCUSSION}
\subsection{The Physical Properties of Radio Plasma}
Double or triple
radio structures similar to NGC~7319 are found in many Seyfert galaxies 
(Ulvestad \& Wilson 1984a, 1984b, 1989; Kukula et al. 1995). 
The steep spectra of the three 
compact components in NGC~7319 are consistent with synchrotron radiation.
Adopting the minimum energy condition for cosmic rays 
plus magnetic field,
we have estimated the magnetic field strength for components A and C.
We have assumed power-law spectra ($\alpha = -1.2$ for A,
$\alpha = -1.3$ for C) between 
10 MHz and 100 GHz and a value
of 100 for the ratio of the total cosmic ray energy 
to the relativistic electron energy.

\par
The results for components A and C are $6\times10^{-4}$ G and 
$5\times10^{-4}$ G, respectively.
The relativistic pressures derived for A and C are $3\times10^{-8}$,
and $2\times10^{-8}$ dyn cm$^{-2}$, respectively.
The thermal gas pressure of the emission-line region of NGC~7319, for
which
the electron density is $n_{e}$=300-600 cm$^{-3}$ (Aoki et al. 1996), 
is $4-8\times10^{-10}$ dyn cm$^{-2}$,
assuming an electron temperature of T$_{e}$ = 10,000 K.
The relativistic pressure is thus at least one or two orders
of magnitude higher than the
thermal gas pressure.

\subsection{Comparison of Radio Data with {\it HST} and Ground-based Spectroscopic Data}
The absolute astrometric uncertainties for {\it HST} data are of the order of 
1\arcsec, so the {\it HST} image cannot be 
registered with the VLA images by relying on the internal
{\it HST} astrometry.
We therefore attempted to obtain more accurate astrometric coordinates
for the {\it HST} image using
the following procedure.
The {\it HST} image was rotated to the cardinal orientation 
by means of the keyword `ORIENTAT' in the 
data header.
Next, the image was smoothed with a gaussian function of 1\farcs5~FWHM
to simulate ground-based seeing.
Various reasonable sizes of the gaussian were tried, but in all
cases the positions of the peak agreed to 0\farcs1.
The peaks in the smoothed and unsmoothed images agree
with each other to 0\farcs08 .
We then assigned the absolute position of the nucleus determined
by ground-based astrometry (Clements 1983) to
the peak of the smoothed image.
\par
However, this registration caused a systematic separation ($\sim$ 0\farcs7)
between features in the {\it HST} and radio images.
The shift of 0\farcs7 is larger than the internal errors of Clements' (1983)
and the VLA astrometry, but there can be uncertainties
in Clements' position due to
structure in the nuclear region and the fact that the spectral sensitivity
of Clements' plates is different to the HST image.
There are also systematic differences between the optical and radio 
astrometric frames.
We finally decided to shift the optical peak onto the position of radio
component
B (Figs 1 and 2), while recognizing the somewhat arbitrary nature of this
alignment and the consequent uncertainty in the interpretation which follows.
The resulting shift of the original {\it HST} coordinates is 1\farcs26.
Using this registration, contours of the 3.6 cm radio image are overlaid on 
the {\it HST} image 
in Figure 6.
Radio component A coincides with the northern arm
of the spiral-like structure and component C is just inside
of the V-shaped feature. Even with the alignment based on Clements'
nuclear position, radio component C is still to the NE of the V-shaped
feature, so we regard this displacement as secure. 
\par
{\it HST} observations have revealed various associations between radio 
continuum and optical line-emission in Seyfert galaxies.
In some Seyfert galaxies, the emission-line features
are shell-like or arc-like and
surround the radio emission;
Mrk~573 is
an excellent example (Capetti et al. 1996; Falcke, Wilson \& Simpson 1998).
This type of 
structure is interpreted as evidence that the ionized gas is 
compressed by the shocks created by the outward motion of the radio plasma.
Such compression increases the
gas density and enhances the surface brightness in line emission.
The relationship between the V-shaped feature and radio
component C is similar to
that between the emission-line arcs and the radio hotspots in Mrk~573.

\par
The emission-lines show a pronounced
blueshift and the line profiles are flat-topped or double-peaked 
near the V-shaped feature (Fig. 7). 
There is also a blueward sloping asymmetry to the [\ion{O}{3}] line 
profiles at the nucleus and northeast of the nucleus (Fig. 7).
 \par
We interpret these observational results as follows. 
The V-shaped feature represents compressed gas behind a bow
shock driven into the photoionized gas by the outwardly moving
radio component C.
This compressed gas is given a motion outward by component C,
and thus shows a large blueshift 
although there is not any independent evidence that the SW is the near side
of the outflow yet.
It is interesting that to the NE of component C, [\ion{N}{2}] 6584 $>$ 
H$\alpha$, while to the SW of component C, H$\alpha$ $>$ [\ion{N}{2}] 6584 (Fig. 7).
There thus appears to be a change of excitation possibly associated
with the bow shock.
H$\alpha$/[\ion{O}{3}], however, does not significantly change across 
the position of component C when we reanalyzed the spectra of 
Aoki et al. (1996).
This may be due to that each emission line was observed on
different nights under different seeing conditions.
More accurate observations are necessary to study change of excitation across
component C.
The blue wing in [\ion{O}{3}] at the nucleus and on the northeastern side 
of the nucleus may be related to the spiral-like structure in the
HST image and/or components A and B in the radio images.
This spiral-like structure could be a curved jet similar to
that seen in ESO 428-G14 
(Falcke et al. 1996) and NGC~4258 (Cecil, Wilson, \& Tully 1992).
\section{CONCLUSIONS}
We have observed the Seyfert galaxy NGC~7319 with the VLA and found
a triple radio source straddling the nucleus.
We have also found a spiral-like structure and
a V-shaped feature in an {\it HST} archival WFPC2 image
of this galaxy.
These optical features are closely related to the radio emissions.
Combining these results with ground-based optical spectroscopy, we
interpret the V-shaped feature as gas compressed by a bow shock driven
into the ambient medium by 
the outwardly moving radio plasmoid. 
To confirm this interpretation, high resolution emission-line imaging 
and spectroscopy are 
needed.

\acknowledgments
We thank the staff of the Array Operation Center in Socorro,
especially G. Taylor, 
for their kind help during observations and data reduction.
We also thank Masaru Watanabe and Hiroshi Ohtani for their valuable 
suggestions and encouragements.
Part of the data analysis was done at the Astronomical Data Analysis Center, 
National Astronomical Observatory of Japan, which is
an inter-university research
institute operated by Ministry of Education, Science, Culture and Sports.
The National Radio Astronomy Observatory is a facility of the National Science Foundation
operated under cooperative agreement by Associated Universities, Inc.
This research was supported in part by NASA through grant NAG 81027.

\newpage

\figcaption{The 3.6 cm naturally weighted image of NGC~7319.
The compact components are labeled A, B, and C.
The beam size is $0\farcs27\times0\farcs26$~FWHM.}

\figcaption{Same as Figure 1, but for the 6 cm naturally weighted image.
The beam size is $0\farcs47\times 0\farcs45$~FWHM.}

\figcaption{Same as Figure 1, but for the 20 cm uniformly weighted image.
The beam size is $1\farcs31\times 1\farcs29$~FWHM.
Components A and B are merged together by the lower resolution than attained
at 3.6 and 6 cm.}

\figcaption{{\it HST} WFPC2 F606W image of NGC~7319.
North is up, and east is to the left. The field of view is $10\arcsec\times10\arcsec$.}

\figcaption{Fine scale structure in the {\it HST} WFPC2 F606W image of 
NGC~7319.
The image was made by
subtracting a smoothed image from the original HST image (see text).
North is up, and east is to the left.
The field of view is $10\arcsec\times 10\arcsec$, so the spatial scale is
the same as Figure 4.}

\figcaption{Contours of the VLA 3.6 cm map are overlaid on the {\it HST}
F606W image. The two images were registered by
assuming that radio component B corresponds
to the peak in the {\it HST} F606W image (see text).}

\figcaption{The variations of line profile along P.A. 27\arcdeg. 
({\it Left}) The `ladder' overlaid on the 3.6 cm VLA 
image of NGC~7319 shows the slit.
Each small rectangle corresponds to a $1\farcs8\times1\farcs46$ slit increment.
({\it Center}) [O III] $\lambda5007$ profiles from Aoki et al. (1996). 
The northeast and southwest ends of the slit are indicated.
The numbers on the left give slit increment numbers. 
The arrows on the right mark positions of the radio components A, B,
the optical 
continuum peak, the radio component C and the V-shaped feature, respectively. 
The spatial and velocity scales are indicated.
The small tick marks indicate 1000 km s$^{-1}$ intervals. 
The long tick mark indicates the systemic velocity of 6740 km s$^{-1}$. 
({\it Right}) Same as ({\it Center}), but 
H$\alpha$+[N II] $\lambda\lambda6548$, 6583 profiles. }

\newpage
\begin{deluxetable}{cccccccccc}
\footnotesize
\tablewidth{0pt}
\tablecolumns{10}
\tablecaption{Parameters of the radio components in NGC~7319}
\tablehead{
\colhead{} & \colhead{$\alpha$\tablenotemark{a}} & \colhead{$\delta$\tablenotemark{a}} & \colhead{3.6 cm} & 
\colhead{3.6 cm} & \colhead{6 cm} & \colhead{6 cm} & \colhead{20 cm} & 
\colhead{Spectral} & \colhead{Spectral} \\ 
\colhead{} & \colhead{J2000} & \colhead{J2000} & \colhead{$f_{\nu}$} & \colhead{Size} & \colhead{$f_{\nu}$} & \colhead{Size} & \colhead{$f_{\nu}$} & 
\colhead{index} & \colhead{index}  \\
\colhead{} & \colhead{(22h36m+)} &\colhead{(33\arcdeg58\arcmin+)} &
\colhead{(mJy)} &
\colhead{(\arcsec)} &
\colhead{(mJy)} &
\colhead{(\arcsec)} &
\colhead{(mJy)} &
\colhead{3.6-6 cm} &
\colhead{6-20 cm} 
}
\startdata
A & 03.604s & 33\farcs95 & 2.1 & 0.3$\times$0.2 & 4.1 & 0.4$\times$0.3 & 19\tablenotemark{b} & -1.2 & -- \nl
B & 03.578s & 33\farcs04 & 0.5 & -- & 1 & -- & -- & -1.4 & -- \nl
C & 03.460s & 30\farcs02 & 0.9 & 0.25$\times$0.2 & 1.8 & 0.5$\times$0.4 & 9 & -1.3 & -1.3 \nl
\enddata
\tablenotetext{a}{These are mean of 3.6 and 6 cm values.}
\tablenotetext{b}{This is the sum of the components A and B.}
\end{deluxetable}
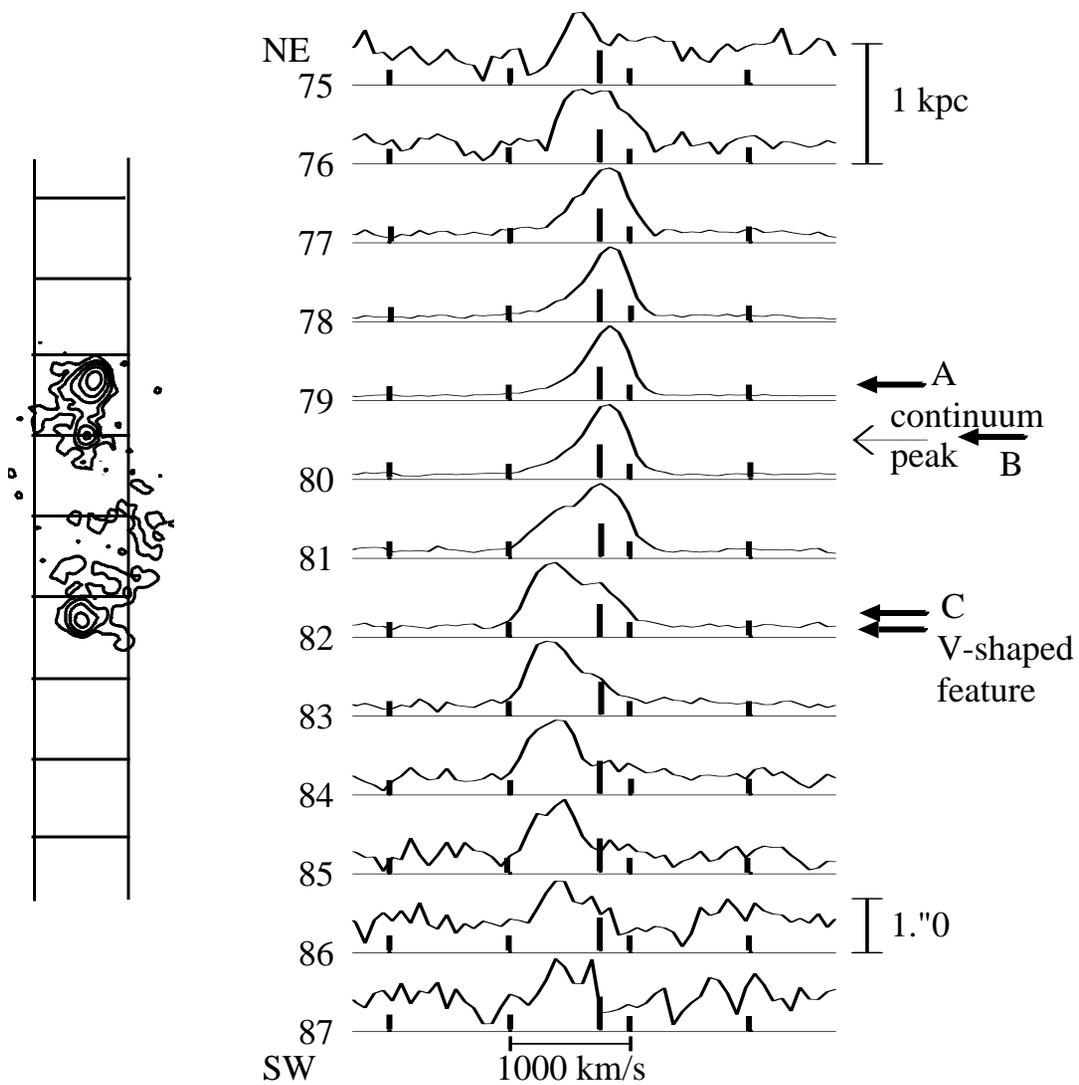
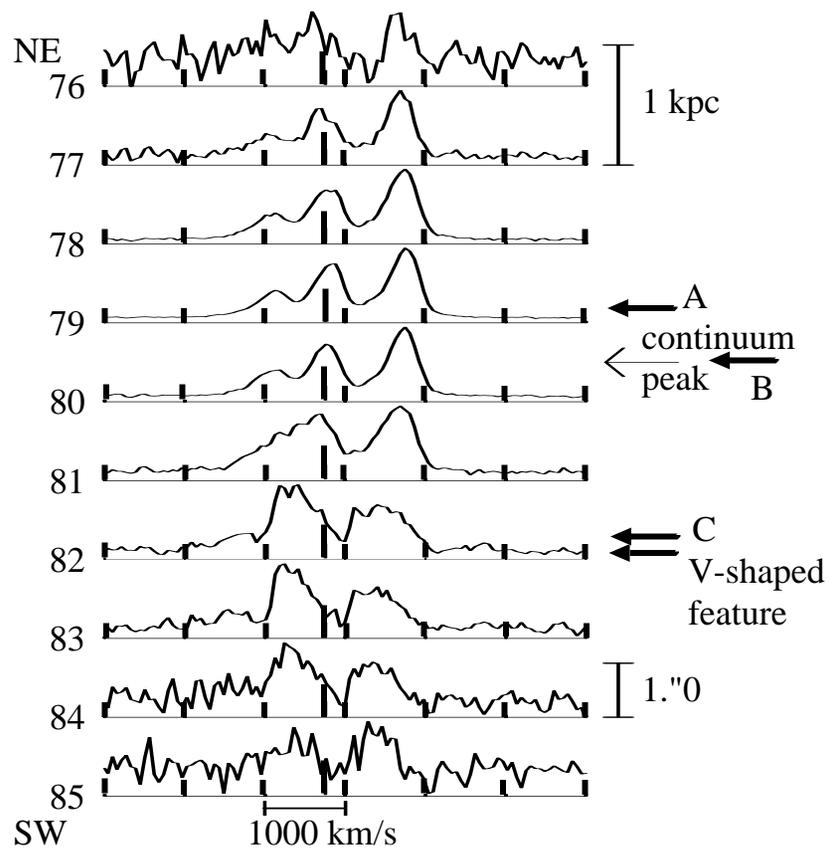
\end{document}